\documentclass{iopjournal}
\usepackage{amssymb}

\begin{document}


\title{Relativity with or without light and Maxwell}

\author{Dragan V. Redžić}

\affil{Faculty of Physics, University of Belgrade, PO
Box 44, 11000 Beograd, Serbia, retired}



\email{redzic@ff.bg.ac.rs}


\begin{abstract}
The complex relationship between Einstein's second postulate and the
Maxwell electromagnetic theory is elucidated. A simple deduction of
the main results of the Ignatowski approach to the theory of
relativity is given. The peculiar status of the principle of
relativity among the Maxwellians is illustrated.
\end{abstract}

\section{Introduction}
Recent papers published in pedagogical journals [1-9] testify that
the problem of how to teach special relativity continues to be a
fascinating topic, discussed at various levels of sophistication.
This is no wonder because the concepts of where, when and why are
essential for our existence as conscious beings, and with the theory
of relativity our instinctive world view is at stake. It is a common
experience of high school physics teachers that even the students
with an aversion to physics lose their apathy and listen with ears
wide open when confronted with the theory of relativity .

The historical path to special relativity \emph{stricto sensu} began
with the second postulate introduced by Einstein in 1905 [10].
Immediately after the publication of reference [10], he noted that
the second postulate is contained in Maxwell's equations [11],
suggesting thus the approach recently elaborated by Aguirregabiria
{\it et al} ([1], [12]).

In the present paper, which was inspired by references [1-12], we
venture to illuminate some dark corners and to simplify some
involved points in relativistic arguments. We believe that the
present attempt could save some time and effort to the teachers and
students of special relativity at undergraduate level, and provides
some fresh insights. In Section 2 we discuss the complex
relationship between Maxwell's equations and the second postulate.
Section 3 is an attempt to deduce the main results of the
`relativity without light' approach in a simple way. The concluding
Section 4 contains a praise of the second postulate and a historical
remark on the peculiar status of the principle of relativity among
the Maxwellians in the late nineteenth century.

\section{Maxwell's equations, `inertia-time' and `light-time' }

Starting from the postulated validity of Maxwell's equations in an
inertial frame $S$ involves that the meaning of `time' has already
been settled in the frame $S$, according to the standard definition
based on Galileo's principle of inertia (`inertia-time'). As is well
known, inertia-time and the validity of Maxwell's equations in the
frame $S$ imply that one can also introduce `light-time' in that
frame, based on rectilinear propagation of electromagnetic
waves\footnote [1] {Following Maxwell (1864), ``light itself [...]
is an electromagnetic disturbance in the form of waves propagated
through the electromagnetic field according to electromagnetic
laws.'' [13]} in vacuum at the same speed $c \equiv
(\epsilon_0\mu_0)^{-1/2}$ in all directions; light-time is identical
with inertia-time in $S$. Applying now the principle of relativity
(assuming also that $\epsilon_0$ and $\mu_0$ are frame-independent),
it follows that one should look for coordinate transformations
between two inertial frames $S$ and $S'$, in uniform motion with
respect to one another, that are consistent with light-time. (While
inertia-time and light-time are identical in any frame, inertia-time
puts no constraints on the transformations sought except that the
validity of Galileo's principle of inertia in $S$ should imply its
validity in $S'$.) Clearly, the transformations cannot be of the
Galilean type, involving $t' = t$, since such transformations
destroy invariance of $c$.\footnote [2] {The fact that `now' is not
the same for all inertial observers is a miracle of different kind
than the one that the motion of the moon about the earth and the
falling of an apple have a common root. (Newton proved the latter in
the third book of {\it Principia}.) According to Newton, ``relative,
apparent and common time is any sensible and external measure
(whether accurate or unequable) of duration by means of motion;''
``duration or the perseverance of the existence of things'' is
Newton's synonym for his ``absolute, true and mathematical time.''
Einstein's postulated uniform propagation of light in vacuum, an
ideal time--keeper (Silberstein's term) identical for all inertial
observers and inaccessible to experimental verification, appears to
be a perfect {\it analogon} of Newton's absolute time which ``of
itself and of its own nature, without relation to anything external,
flows equably,'' and the parts of which ``make no impressions on the
senses'' (``{\it non incurrunt in sensus};'' all quotations of
Newton are from the first {\it Scholium} of {\it Principia} [14]).
Newton, and all physicists before Einstein (including Voigt, Larmor,
Lorentz and Poincar\' e \cite {WR}-\cite {HB}), took it for granted
that there was only one `time,' absolute Newtonian time, for all
observers in motion with respect to one another. Einstein was bold
enough to venture that each inertial observer has her/his own {\it
absolute Einsteinian time}.} Assumed uniformity of space and time
implies that the transformations must be linear; employing also
invariance of the speed of light $c$ and assuming isotropy of space,
one obtains the Lorentz transformations.\footnote [3] {Einstein's
original derivation of the Lorentz transformations [10], while
cumbersome, is perfectly correct, without involving {\it Galilean
transformations}, as clarified by Mart\' inez [19]. Einstein for the
first time gave the explanation of how uniformity of space and time
implies the linearity of the transformations in [20], Section 7.4;
in the same paper he gave for the first time a definition of a clock
([20], p 21), that involves ``the principle of sufficient reason.''}

The above argument seems to be condensed in the footnote from
Einstein's second relativity paper: ``The principle of the constancy
of the velocity of light used there [in [10]] is of course contained
in Maxwell's equations.'' [11] The laconic footnote probably reveals
Einstein's {\it original} train of thought leading him to special
relativity: from Maxwell's equations to the constancy of the speed
of light, and through the principle of relativity and properties of
space and time, to the Lorentz transformations.

However, in 1905 instead of starting from Maxwell's equations
Einstein chose a different path [10]. Why? He gave an explanation
thirty years later, at the beginning of his penultimate published
attempt to derive the mass--energy relation: ``The special theory of
relativity grew out of the Maxwell electromagnetic equations. So it
came about that even in the derivation of the mechanical concepts
and their relations the consideration of those of the
electromagnetic field has played an essential role. The question as
to the independence of those relations is a natural one because the
Lorentz transformation, the real basis of the special relativity
theory, in itself has nothing to do with the Maxwell theory and
because we do not know the extent to which the energy concepts of
the Maxwell theory can be maintained in the face of the data of
molecular physics.'' [21]

However, as noted in [22], Einstein in 1905 would have had to derive
the Lorentz transformations following the template of
thermodynamics, making them independent of the Maxwell theory, even
if he had been completely ignorant of light quanta or Planck's 1900
derivation of the black--body radiation formula. Namely, a
definition of time that is as simple as possible must conceptually
precede any discussion about ``the laws according to which the
states of physical systems change'' (``die Gesetze nach denen sich
die Zust\"{a}nde der physikalischen Systeme \"{a}ndern'' [10]),
including the domain of the validity of Maxwell's equations.
Attempting to keep the Lorentz transformations divorced from the
Maxwell electromagnetic theory, Einstein took a necessary condition
of the Maxwell theory, replacing ``electromagnetic waves'' with
``light,'' to be his principle of the constancy of the speed of
light (the notorious ``second postulate''). This principle, which
appears to be `simplicity itself,' combined with the
(meta-)principle of relativity, is the crux of special relativity.
Einstein's `definite velocity $V$' of propagation of light in vacuum
(he used the symbol `$V$' instead of `$c$' in [10]), appearing in
the introduction and in paragraph 2 of [10], does not have the
familiar meaning of a derived quantity,  $V =$ (light path)/time
interval, where `time interval' is a segment of inertia--time.
Instead, Einstein's $V$ is a primitive quantity (of course, all
other speeds are derived quantities), and `time interval' is {\it by
definition} equal to (one--way light path)/$V$. Thus the speed of
light in vacuum is by definition equal to the universal constant
$V$, identical for all inertial observers, in perfect disagreement
with our instinctive Galilean mentality.\footnote [4] {As is
recalled in [23], Einstein's second postulate has a rather intricate
content. It postulates that, relative to an inertial frame, ``the
one-clock two-way speed of light in vacuum $V$ (a measurable
physical quantity and, as measurements reveal, a universal constant)
is constant and independent of the velocity of light source and
equals the one-way two-clock speed of light in vacuum (an
immeasurable quantity). [...] It should be pointed out that
throughout the Relativity Paper Einstein used the same symbol
(`$V$') for the speed of light in vacuum and the phase velocity of
electromagnetic waves in vacuum according to the `Maxwell-Hertz
equations', linking thus special relativity with Maxwell's theory,
and at the same time linking the new time keeper (propagation of
light) with the earlier ones (cf [24]).''} Two circumstances were
decisive for Einstein's argument: first, Maxwell's equations can be
simply made Lorentz--covariant and thus consistent with the
principle of relativity by appropriately defining the
transformations for the electric and magnetic fields and charge and
current densities (the fact already known (partly) to Lorentz [25]
and (fully) to Poincar\' e [26], but also (partly) to Larmor [27],
cf [16]), and, second, empirical evidence for Maxwell's theory in
pseudo--inertial frame tied to the earth.

The above discussion reveals that Einstein's second postulate
conceptually precedes the principle of relativity. The rich content
of the second postulate is understated in the laconic Einstein's
footnote [11] quoted above.

\section{Relativity without the light postulate }

The traditional way of deriving the Lorentz transformations,
starting from Einstein [10], is {\it basically} based on the
following assumptions:

1. There exists a frame of reference in which spatial coordinates
$x$, $y$ and $z$ are Cartesian coordinates (Euclidean geometry
applies), and time coordinate $t$ is defined through `light--time'
(light signals in vacuum propagate rectilinearly and uniformly and
the one--way two--clock speed of light is by definition the known
constant, $c \equiv (\epsilon_0\mu_0)^{-1/2}$); we will call a frame
of reference with such properties of space and time the `primitive
frame.'

2. Each observer in uniform motion with respect to the primitive
frame (`primitive observer') has her/his own primitive frame so the
speed of light is the universal constant, $c$, the same for all
primitive observers.

3. The principle of relativity applies to primitive frames.

4. Uniformity of space and time and isotropy of space.

The above assumptions are sufficient for deriving the Lorentz
transformations. Since light--time is consistent with inertia--time,
by adding Galileo's principle of inertia each primitive frame is
also inertial frame, but Galileo's principle is not necessary for
deriving the Lorentz transformations.

One may ask at which coordinate transformations one arrives with the
principle of relativity but without light and the second postulate,
replacing in the above assumptions primitive frame with inertial
frame, employing inertia--time based on Galileo's principle of
inertia. As is well known, we owe the query and the correct answer
to it to Vladimir Ignatowski [28]. In what follows we give a simple
derivation of Ignatowski's main results.

Consider two inertial frames $S$ and $S'$ in the standard
configuration ($S'$ is in uniform motion with respect to $S$ along
the common positive $x-x'$ axes at speed $v$, $y-$ and $z-$axis
parallel to the $y'-$ and $z'-$axis, respectively, the origins
coincide at $t = t' = 0$), Uniformity of space and time implies that
required coordinate transformations must be linear; also, isotropy
of space implies that $x'$ is independent of $y$ and $z$.
Consequently,

\begin{equation}
x' = x/F_v + t/G_v\, ,
\end{equation}
where $F_v$ and $G_v$ are parameters dependent solely on $v$, as yet
unknown. Setting for simplicity $y' = y$ and $z' = z$ (for a
convincing \emph{deduction} see [10]), our considerations will be
restricted to $x$ and $t$ only. Employing the principle of
relativity, {\it assuming} velocity reciprocity\footnote [5]
{Velocity reciprocity means that the velocity of the $S'$ frame with
respect to the $S$ frame is the opposite of the velocity of $S$ with
respect to $S'$. Many an author, including Ignatowski [28], takes
velocity reciprocity as an immediate and self-evident consequence
only of the principle of relativity. However, various additional
assumptions are also required, such as uniformity of space and of
time, spatial and temporal isotropy, and causality, as pointed out
by Berzi and Gorini [29] and by L\' evy-Leblond [30]; a thorough
discussion of the issue was recently published by Patrick Moylan
[5].

By the way, Einstein's first derivation of the Lorentz
transformations (given in paragraph 3 of [10]) based on the
postulated, finite and known, universal speed $V$ (Einstein's symbol
for `$c$'), does not assume velocity reciprocity, but \emph{deduces
}it. It is perhaps amusing to note that therein he denotes by
$\varphi(v)$ both his $a$ and $a\beta$, where $\beta \equiv
\frac{1}{\sqrt {1 - (v/V)^2}}$. Moreover, he employs three reference
frames (which he calls `coordinate systems'), `resting frame' $K$
with coordinates $x,y,z$ and $t$, `moving frame' $k$ with
coordinates $\xi,\eta,\zeta$ and $\tau$, and a third frame $K'$ with
coordinates $x',y',z'$ and $t'$, moving relative to $k$. In the same
paragraph 3 of [10] he uses the symbol $x'$ also for $x - vt$, so
$x',y,z$ and $t$ refer to {\it coordinate systems} introduced in the
{\it reference frame} $K$. All that of course does not add to
lucidity; reading Einstein, like reading Maxwell, is always an
adventure {\it par excellence} [31,32].}, one has

\begin{equation}
x = x'/F_{-v} + t'/G_{-v}\,
\end{equation}
Considering trajectories of the origins and lengths of moving unit
rods, one obtains

\begin{equation}
1/G_v = -v/F_v \, ,\quad 1/G_{-v} = v/F_{-v}\, , \quad F_v =
F_{-v}\, .
\end{equation}
Eqs. (1)-(3) give [33]

\begin{equation}
x' = \frac {x - vt}{F_v}\, ,  \quad  \mbox {and} \quad t' = \frac {t
- \kappa_v x}{F_v}\, ,
\end{equation}
where

\begin{equation}
\kappa_v \equiv \frac {1 - F^2_v}{v}\, .
\end{equation}

It is convenient to write eqs. (4) in matrix form [34,35]

\begin{equation}
\left( \begin{array}{c}
t' \\
x'
\end{array} \right )
 = \frac {1}{F_v} \left( \begin{array}{cc}
1 & -\kappa_v\\
-v & 1
\end{array} \right ) \left( \begin{array}{c}
t \\
x
\end{array} \right ) \equiv \Lambda_v \left( \begin{array}{c}
t \\
x
\end{array} \right )
\end{equation}

Introduce now a third inertial frame $S''$, in the standard
configuration with $S'$, moving at the speed $u$ relative to $S'$.
One has

\begin{equation}
\left( \begin{array}{c}
t'' \\
x''
\end{array} \right )
 = \frac {1}{F_u} \left( \begin{array}{cc}
1 & -\kappa_u\\
-u & 1
\end{array} \right ) \left( \begin{array}{c}
t' \\
x'
\end{array} \right ) \equiv \Lambda_u \left( \begin{array}{c}
t' \\
x'
\end{array} \right )
\end{equation}
where $\kappa_u \equiv (1 - F^2_u)/u$. Eqs. (6) and (7) obviously
give

\begin{equation}
\left( \begin{array}{c}
t'' \\
x''
\end{array} \right )
 = \frac {1}{F_u F_v} \left( \begin{array}{cc}
1 + \kappa_u v & -\kappa_u -\kappa_v\\
-u - v& 1 + \kappa_v u
\end{array} \right ) \left( \begin{array}{c}
t \\
x
\end{array} \right ) \equiv \Lambda_u \Lambda_v\left( \begin{array}{c}
t \\
x
\end{array} \right )
\end{equation}

It is natural to require that coordinate transformations between two
inertial frames satisfy closure property, i.e. the successive
application of two such transformations yields a third such
transformation of the same form and content. Comparing eqs. (8) and
(6), the requirement can be satisfied if and only if

\begin{equation}
\kappa_u v = \kappa_v u \, ,
\end{equation}
{\it and}

\begin{equation}
1 + \kappa_u v \neq 0 \, .
\end{equation}

Condition (9) is tantamount to

\begin{equation}
\frac {\kappa_u}{u} =  \frac {\kappa_v}{v} \equiv \Omega \, .
\end{equation}
Since $u$ and $v$ are arbitrary, $\Omega$ must be a universal
constant, the same for all inertial observers. Conditions (10) and
(11) give

\begin{equation}
1 + \Omega u v \neq 0 \, ,
\end{equation}
which excludes the possibility of negative $\Omega$ (otherwise, one
would have $1 + \Omega u v = 0$  for $uv = 1/|\Omega|$). Thus the
universal constant $\Omega$ must be nonnegative,

\begin{equation}
\Omega \geq 0 \, .
\end{equation}

Expressed in terms of $\Omega$, employing condition (12), matrices
$\Lambda_v$ and $\Lambda_u \Lambda_v$ become

\begin{equation}
\Lambda_v \equiv \frac {1}{F_v} \left( \begin{array}{cc}
1 & - v \Omega\\
-v & 1
\end{array} \right )\, ,
\end{equation}

\begin{equation}
\Lambda_u \Lambda_v \equiv \frac {1}{F_w} \left( \begin{array}{cc}
1 & - w \Omega\\
-w & 1
\end{array} \right )\, ,
\end{equation}
where

\begin{equation}
w \equiv \frac {v + u}{1 + \Omega v u}\, ,
\end{equation}
and

\begin{equation}
 \frac {1}{F_w}\equiv \frac {1 + \Omega u v}{F_u F_v}\, .
\end{equation}
Equations (5) and (11) imply

\begin{equation}
F_v =  \sqrt{1 - \Omega v^2}\, ,
\end{equation}
since only the positive value of $F_v$ makes sense. Thus also $F_u =
\sqrt{1 - \Omega u^2}$, and $F_w$ from eq. (17) must be equal to
$\sqrt{1 - \Omega w^2}$ with $w$ given by eq. (16). One can verify
that this is indeed so and consequently $w$ is the speed of the
frame $S''$ with respect to the $S$ frame.

\newpage

There are two possible choices for $\Omega$, zero or a positive
value. $\Omega = 0$ yields the Galilean transformations

\begin{equation}
x' = x-vt\, ,\quad  y' = y\, ,\quad  z' = z\, , \quad t' = t\, ,
\end{equation}
which imply that there is one time for all inertial observers
(identical with Newton's absolute time), that a standard of length
always has the same length independent of its velocity, and that $v$
can be arbitrarily large. While in perfect agreement with our
instinctive Galilean mentality, the Galilean transformations cannot
be made compatible with a consistent interpretation of
experience.\footnote [6] {Newton's absolute space (that ``by its own
nature without relation to any thing external remains always uniform
and unchangeable'') and absolute time (that ``by its own nature
without relation to any thing external flows equably'') appear to be
the natural habitat for the Galilean transformations. As the parts
of absolute space ``do by no means come under the observation of our
senses'' (``{\it non incurrunt in sensus}''), and the same applies
to absolute time, it appears that those {\it absolute} concepts
remain transcendental, in Kantian sense (compare Maxwell's poetic
discussion in his {\it Matter and Motion} [36]; by the way, what
Maxwell calls ``the doctrine of relativity of all physical
phenomena'' is the root of Poincar\' e--Einstein's principle of
relativity). But, keeping in mind that ``everything which is not
forbidden is allowed,'' one should not dispense with a possible
interpretation of the Galilean transformations.

It should be stressed that in special relativity all inertial frames
share one and the same space, cf. Section 3 of [10], which is
analogous to Newton's absolute space, without of course involving
the idea of absolute rest.}

The choice $\Omega > 0$ yields the well-known Ignatowski
transformations [28]

\begin{equation}
x' = \frac {1}{\sqrt{1 - \Omega v^2}} (x-vt)\, ,\quad  y' = y\,
,\quad z' = z\, , \quad t' = \frac {1}{\sqrt{1 - \Omega v^2}} (t -
\Omega v x)\, .
\end{equation}
Putting

\begin{equation}
\overline c \equiv  \frac {1}{\sqrt{\Omega }} \, ,
\end{equation}
and employing eq. (18), one obtains that

\begin{equation}
v < \overline c\, .
\end{equation}
Thus $\overline c$ is the universal {\it limit} speed for massive
particles, as yet unknown. Putting also

\begin{equation}
\overline \gamma_v \equiv \frac {1}{\sqrt{1 - v^2/\overline c^2}}\,
.
\end{equation}
eqs. (20), (16) and (17) become

\begin{equation}
x' = \overline \gamma_v (x-vt)\, ,\quad  y' = y\, ,\quad  z' = z\, ,
\quad t' = \overline \gamma_v (t - vx/\overline c^2)\, ,
\end{equation}

\begin{equation}
w \equiv \frac {v + u}{1 + vu/\overline c^2}\, ,
\end{equation}

\begin{equation}
\overline \gamma_w \equiv \overline \gamma_u \overline \gamma_v (1 +
vu/\overline c^2)\, .
\end{equation}


Since we concluded above that $w$ is the speed of $S''$ relative to
$S$, in eq. (25) one recognizes the familiar relativistic
composition of velocities, and in eq. (26) the familiar relativistic
transformation for gamma factors (cf, e. g., [37, 38]). It can be
shown, employing eqs. (24) and invariance of causality, that the
speed of propagation of interaction must be less than or equal to
$\overline c$, but the theory does not specify whether $\overline c$
is a physically attainable speed or not \cite {WR1}.

Thus the Ignatowski transformations (20) are Lorentz--like
transformations. To obtain the Lorentz transformations, i.e. to
ascertain the value of the universal constant $\overline c$, one
must turn to phenomena, without invoking properties of light or
Maxwell's electromagnetic theory.

\section{Concluding comments}

Turning to phenomena, however, is tricky considering that inertial
observers inhabit ``ideal infinitely extended gravity-free inertial
frames,'' whereas in the real world their habitat is restricted to
``the freely falling nonrotating local frames'' [37].\footnote [7]
{This fact was {\it basically} well known to Newton, see corollaries
V and VI of the Laws of Motion [14], and also [39].} We quoted above
Einstein's clear-cut statement that ``the Lorentz transformation,
the real basis of the special relativity theory, in itself has
nothing to do with the Maxwell theory [...],'' which is widely
endorsed in the literature. (``Special relativity would exist even
if light and electromagnetism were somehow eliminated from
nature.''[37]) While various scenarios that have been proposed to
ascertain the value of $\overline c$, without invoking properties of
light or Maxwell's electromagnetic theory, appear to be free from
logical errors (cf, e. g., [6, 33, 40]), none of them can be
actually implemented.

On the other hand, Einstein's second postulate combined with the
principle of relativity states, basically: first, there is a
universal finite speed, c, the same for all {\it primitive
observers}, and light propagates in vacuum at the universal speed;
second, the universal speed c is a primitive quantity, and `time'
(`light--time', `duration') in any primitive frame is a derived
quantity through time interval $\stackrel{\mbox{\small d}}{=}$ light
path/c; third, the magnitude of $c$ is equal to the speed of
propagation of electromagnetic waves in vacuum,
$(\epsilon_0\mu_0)^{-1/2}$ as given by Maxwell's electromagnetic
theory, and also it is equal to the one-clock two-way speed of light
as ascertained by terrestrial measurements and consistent with
astronomical observations.

Thus, whereas in `relativity without light and electromagnetism'
$\overline c$ is the universal limit speed determinable only in
principle through Gedankenexperiments, in the second postulate
approach $c$ is the speed which pertains to a real phenomenon (the
essential one for our experience of the world), ``concerning which
we know something certain [...] in a higher degree than for any
other process which could be considered, thanks to the
investigations of Maxwell and H. A. Lorentz'' [41]. The second
postulate combined with the principle of relativity, while
counterintuitive, gives a simple path to the concepts of time and
primitive frame, and to the Lorentz transformations.\footnote [8]
{It should be stressed that relativistic length contraction and
clock retardation cannot be verified directly. As Jefimenko [42]
notes, Einstein's method for measuring the length of a moving rod
proposed in [10], ``was, of course, merely a `Gedankenexperiment,'
that is, an imaginary procedure, a verbalization of [equation $x' =
\gamma (x-vt)$], that cannot be actually implemented.'' Also, some
familiar relativistic generalizations concerning clock retardation
have to be amended \cite {DVRS}.} The second postulate is more
fundamental than the principle of relativity because a definition of
time that is as simple as possible must conceptually precede any
discussion about ``the laws according to which the states of
physical systems change.''

Is it indeed a surprise that one can deduce the Lorentz--like
transformations from the principle of relativity and Galileo's
principle of inertia, assuming uniformity of space and time, spatial
and temporal isotropy, and causality, without postulating a
universal finite speed, as Einstein did? With hindsight, {\it post
festum ex post facto}, it appears that the Ignatowski's result could
have been anticipated by the following simple argument. If there is
the universal (signal) speed that is infinite, then $t' = t$,
according to clock synchronization. Applying {\it modus tollens}, if
$t' \neq t$, then either the universal speed is finite, or there is
no universal limit speed, the same for all inertial observers (which
possibility contradicts the principle of relativity). Thus the
Ignatowski's result is latent in the seemingly innocent starting
assumption $t' \neq t$. And what could be a more natural candidate
for the universal limit speed in the world of \emph{phenomena} than
the speed of light?

To summarize, Einstein's definition of the ``time'' [10] is
basically based on a tacit assumption that there is a universal
finite speed of propagation of interaction; the limit speed is a
primitive quantity equal to the speed of light from Maxwell's
electromagnetic theory, $c$, and ``time interval'' is a derived
quantity. On the other hand, in Ignatowski's approach, the universal
limit speed of propagation of interaction (whether attainable or
not), $\overline c$, arises as a consequence of his starting
assumptions; the value of the limit speed $\overline c$ remains
unknown, perhaps transcendent, independent of any physical theory.
In this sense Ignatowski's relativity is more general than Eistein's
relativity [44], true, the meaning of Ignatowski's ``$t$'' is
somewhat blurred. Obviously, if $\overline c \neq c$, then Maxwell's
equations are not Lorentz-like covariant, and thus do not conform to
Ignatowski's relativity.

Finally, a historical remark on the peculiar relationship between
the principle of relativity and Maxwell's electromagnetic theory at
the end of the nineteenth century. In their recent papers, Browne
[7] and Moylan [5] discuss briefly the evolution of the principle of
relativity, recalling contributions of the saints and martyrs of the
philosophical calendar, pride of place probably belonging to Galileo
[7,5,44]. Illustrating the crucial role played by Poincar\' e in the
vigorous defense of the relativity principle, Moylan notes that ``at
the end of the nineteenth century, physics was in a terrible state
of confusion. Maxwell's equations were not preserved under the
Galilean transformation [...], the followers of Maxwell's
electrodynamics were ready to uproot the relativity principle and
reinstate a new form of geocentrism, where the relativity principle
no longer held true.'' However, there is a little-known episode that
testifies that the situation was more complex. A topic of discussion
among physicists in the late nineteenth century was the
electrodynamic interaction  between a charge and a current--carrying
loop at relative rest, that are moving uniformly with respect to the
ether [45-47], cf also [48,49]. The basic feeling of the `old'
physicists was simple: it is highly improbable that anything depends
on the motion with respect to the ether; physical effects depend
only on the relative motion between ponderable bodies and on their
mutual relative position. Since in the problem considered Maxwell's
theory classically interpreted predicts a nonzero force [50],
depending on unobservable speed $v$ of the system with respect to
the ether, Budde [45], FitzGerald [46] and Lorentz [47] postulated
that charges were induced on the current loop in exactly that amount
required to cancel the electrodynamic force due to the motion with
respect to the ether. Their solution to the problem, reached through
a rather contrived Budde's `principle of neutralizing charge' (cf
[48]), up to the second order terms in $v/c$ coincides with what we
think today to be the correct solution. Thus, some of the `old'
physicists were more ready to introduce an {\it ad hoc} hypothesis
than to sacrifice the principle of relativity. Poincar\' e's
melancholy phrase, ``les hypoth\` eses, c'est le fonds qui manque le
moins,''\footnote [9] {Probably echoing the first two verses of the
fable by Jean de La Fontaine, Le Laboureur et ses enfants: {\it
Travaillez, prenez de la peine: C'est le fonds qui manque le
moins.}} expresses his longing for a principle, instead of piling up
hypotheses. The longed-for principle, while counterintuitive and far
from being clearly stated, was the second postulate.

\vspace{5mm}

\noindent \textbf{Funding}

\vspace{2mm}

This research received no external funding.

\vspace {5mm}

\noindent \textbf{Data Availability Statement }

\vspace{2mm}

No new data were generated or analyzed in support of this research.

\vspace{5mm}

\noindent \textbf{Conflicts of interest}

\vspace{2mm}

The author declares no conflicts of interest.

\end{document}